\documentclass[a4paper]{jpconf}
\usepackage{graphicx}
\begin{document}
  \newcommand {\nc} {\newcommand}
  \nc {\beq} {\begin{eqnarray}}
  \nc {\eeq} {\nonumber \end{eqnarray}}
  \nc {\eeqn}[1] {\label {#1} \end{eqnarray}}
  \nc {\Sec} [1] {Sec.~\ref{#1}}
  \nc {\tbl} [1] {Table~\ref{#1}}
  \nc {\fig} [1] {Fig.~\ref{#1}}
  \nc {\Eq} [1] {Eq.(\ref{#1})}
  \nc {\Ref} [1] {Ref.~\cite{#1}}
  \nc {\eq} [1] {(\ref{#1})}
  \nc {\ex} [1] {$^{#1}$}
  \nc {\ve} [1] {\mbox{\boldmath $#1$}}
  \nc {\flim} [2] {\mathop{\longrightarrow}\limits_{{#1}\rightarrow{#2}}}

\title{Study of clustering structures through breakup reactions}

\author{Pierre Capel}

\address{Physique Nucl\'eaire et Physique Quantique (CP229), Universit\'e Libre de Bruxelles (ULB)\\
50 av. F.\ D.\ Roosevelt, B-1050 Brussels, Belgium}

\ead{pierre.capel@ulb.ac.be}

\begin{abstract}
Models for the description of breakup reactions used to study the structure of exotic cluster structures like halos are reviewed.
The sensitivity of these models to the projectile description is presented.
Calculations are sensitive to the projectile ground state mostly through its asymptotic normalisation coefficient (ANC).
They also probe the continuum of the projectile.
This enables studying not only the bound states of the projectile but also its continuum, both resonant and non-resonant.
This opens the possibility to study correlations between both halo neutrons in two-neutron halo nuclei.
\end{abstract}

\section{Introduction}
The developments of radioactive-ion beams in the mid-80s, has permitted the exploration of the nuclear chart far from stability.
Various unexpected features have been uncovered thanks to this technical breakthrough.
Halo nuclei are of particular interest here as they exhibit a strongly clusterised structure that explains there very large radius compared to their isobars.
Indeed these neutron-rich nuclei have a small binding energy for one or two neutrons.
Thanks to this lose binding, the valence neutron(s) can tunnel far into the classically forbidden region to form a dilute halo around a more compact core \cite{Tan96}.
\ex{11}Be and \ex{15}C are examples of one-neutron halo nuclei,
while \ex{6}He and \ex{11}Li are archetypical two-neutron halo nuclei.
Though less probable, proton halos are also possible.
For example, \ex{8}B is thought to exhibit a one-proton halo.

Being located close to the dripline, halo nuclei are very short lived and cannot therefore be studied through usual spectroscopic techniques.
To study their structure, one must resort to indirect techniques such as reactions \cite{Tan96}.
Breakup reaction is probably among the best tools to study exotic clusterised structures such as halo nuclei.
During this reaction, the nucleus of interest is broken up into its constituents---the halo nucleon(s) and the core in this case---through its interaction with a target \cite{Fuk04}.
To extract valuable information about the cluster structure of the projectile from breakup observables, a good understanding of the reaction mechanism is needed and a clear knowledge of the influence of the projectile structure upon the measured cross sections is required.
In this contribution, I will present the nonperturbative reaction models mostly used to analyse breakup data: the Continuum Discretised Coupled Channel model (CDCC), the Time-Dependent approach (TD) and the Dynamical Eikonal Approximation (DEA) (see also Ref.~\cite{BC12} for a recent review).
I will then discuss recent results that show the influence of the projectile description upon breakup data.
An outlook concludes these proceedings.

\section{Breakup modelling}\label{models}
\subsection{Theoretical framework}
In most breakup models, the projectile $P$ is described as a two-body system: an inert core $c$ to which the halo nucleon $f$ is loosely bound.
The $c$-$f$ interaction is described by a phenomenological potential $V_{cf}$ that is adjusted to reproduce the energy and the quantum numbers of the bound states and maybe some resonant states of the nucleus.
The Hamiltonian that describes this cluster structure hence reads
\beq
H_0=T_r+V_{cf}(\ve{r}),
\eeqn{e1}
where $\ve{r}$ is the $c$-$f$ relative coordinate. 
The eigenstates of $H_0$ are the wave functions describing the projectile internal structure
\beq
H_0 \Phi_{lm}(\epsilon,\ve r)=\epsilon\Phi_{lm}(\epsilon,\ve r),
\eeqn{e2}
where $l$ is the $c$-$f$ relative angular momentum and $m$ is its projection (the spins of the clusters are neglected here for simplicity).
The negative-energy eigenstates correspond to the bound states of the system.
In the following the ground state of energy $\epsilon_0$ is denoted by $\Phi_0$.
The positive-energy states describe the $c$-$f$ continuum, i.e.\ the broken up projectile.


The target $T$ is usually described as a structureless particle and its interaction with the projectile constituents $c$ and $f$ is simulated by the optical potentials $V_{cT}$ and $V_{fT}$, respectively.
Within this framework, studying the $P$-$T$ collision reduces to solving the following three-body Schr\"odinger equation
\beq
\left[T_R+H_0+V_{cT}(\ve{R}_{cT})+V_{fT}(\ve{R}_{fT})\right]\Psi(\ve{r},\ve{R})=E_T\Psi(\ve{r},\ve{R}),
\eeqn{e3}
where $\ve R$ is the coordinate of the projectile centre of mass relative to the target, 
while $\ve{R}_{cT}$ and $\ve{R}_{fT}$ are the $c$-$T$ and $f$-$T$ relative coordinates, respectively.
\Eq{e3} is solved with the condition that the projectile, initially in its ground state, is impinging on the target:
\beq
\Psi(\ve{r},\ve{R})\flim{Z}{-\infty}e^{iKZ+\cdots}\Phi_0(\ve{r}),
\eeqn{e5}
where $K$ is the initial $P$-$T$ wave number, which is related to the total energy $E_T=\hbar^2K^2/2\mu_{PT}+\epsilon_0$, with $\mu_{PT}$ the projectile-target reduced mass.

\subsection{CDCC}
In the Continuum Discretised Coupled Channel method (CDCC), \Eq{e3} is solved expanding the three-body wave function $\Psi$ upon the projectile eigenstates $\Phi_{lm}(\epsilon)$ \cite{Kam86,TNT01}
\beq
\Psi(\ve{r},\ve{R})=\sum_i\chi_i(\ve{R})\Phi_i(\ve{r}),
\eeqn{e6}
where $i$ stands for $l$, $m$ and $\epsilon$.
This leads to a set of coupled equations
\beq
\left[T_R+\epsilon+V_{ii}\right]\chi_i+\sum_{j\ne i}V_{ij}\chi_j=E_T\chi_i,
\eeqn{e7}
where $V_{ij}=\langle\Phi_i|V_{cT}+V_{fT}|\Phi_j\rangle$ are coupling the various channels.

Since the model aims at describing the breakup of the projectile, the expansion \eq{e6} must include a reliable description of the continuum of the projectile.
To be tractable, this description is obtained by discretisation of the continuum \cite{Yah86}.

\subsection{Time-dependent approach}
In the Time-Dependent approach (TD) the three-body Schr\"odinger equation \eq{e3} is simplified using a semiclassical approximation, viz. approaching the $P$-$T$ relative motion by a classical trajectory \cite{KYS94,CBM03c}.
Along that trajectory, the projectile experiences a time-dependent potential that simulates its interaction with the target.
This leads to the resolution of a time-dependent Schr\"odinger equation
\beq
i\hbar \frac{\partial}{\partial t}\Psi(\ve{r},\ve{b},t)=[H_0 + V_{cT}(t)+V_{fT}(t)] \Psi(\ve{r},\ve{b},t),
\eeqn{e8}
which is solved with the initial condition $\Psi(\ve{r},\ve{b},t)\flim{t}{-\infty}\Phi_0(\ve r)$ in agreement with \Eq{e5}.

\subsection{Dynamical Eikonal Approximation}
The Dynamical Eikonal Approximation (DEA) is based on the eikonal approximation \cite{Glauber}, which assumes that at sufficiently high beam energy the $P$-$T$ relative motion does not deviate much from the incoming plane wave \eq{e5}.
The idea is thus to factorise that plane wave out of $\Psi$
\beq
\Psi(\ve r,\ve R)=e^{iKZ}\widehat\Psi(\ve r,\ve R)
\eeqn{e9}
to obtain a function $\widehat \Psi$ that varies smoothly with $\ve R$.
Neglecting its second-order derivatives with respect to its first-order derivatives leads to the DEA equation \cite{BCG05}
\beq
i\hbar v \frac{\partial}{\partial Z}\widehat{\Psi}(\ve{r},\ve{b},Z)=[H_0-\epsilon_0+V_{cT}+V_{fT}]
\widehat{\Psi}(\ve{r},\ve{b},Z),
\eeqn{e10}
which is solved with the boundary condition $\widehat\Psi(\ve{r},\ve{R})\flim{Z}{-\infty}\Phi_0(\ve r)$ according to Eqs.~\eq{e5} and\eq{e9}.

\begin{figure}
\center
\includegraphics[trim=1.5cm 18cm 7cm 1.7cm, clip=true, width=8.3cm]{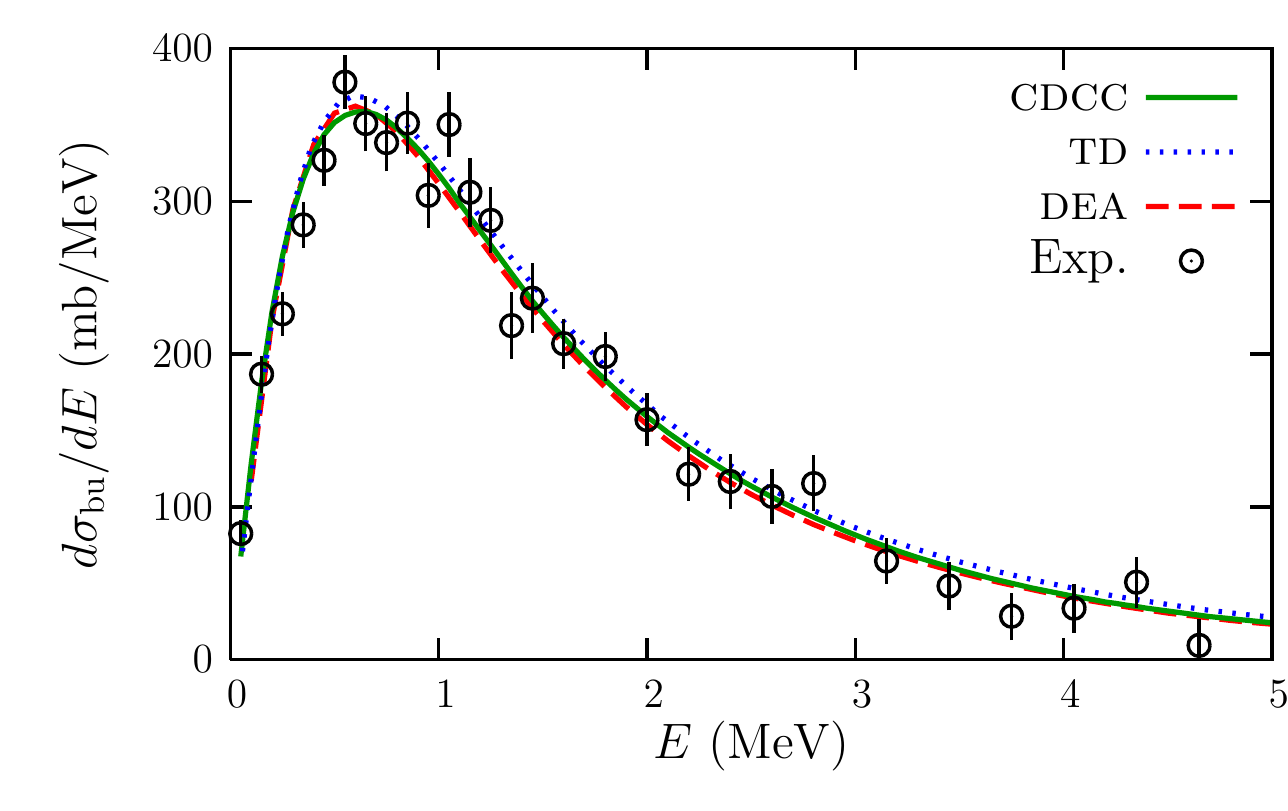}
\includegraphics[trim=2.5cm 18cm 7cm 1.7cm, clip=true, width=7.7cm]{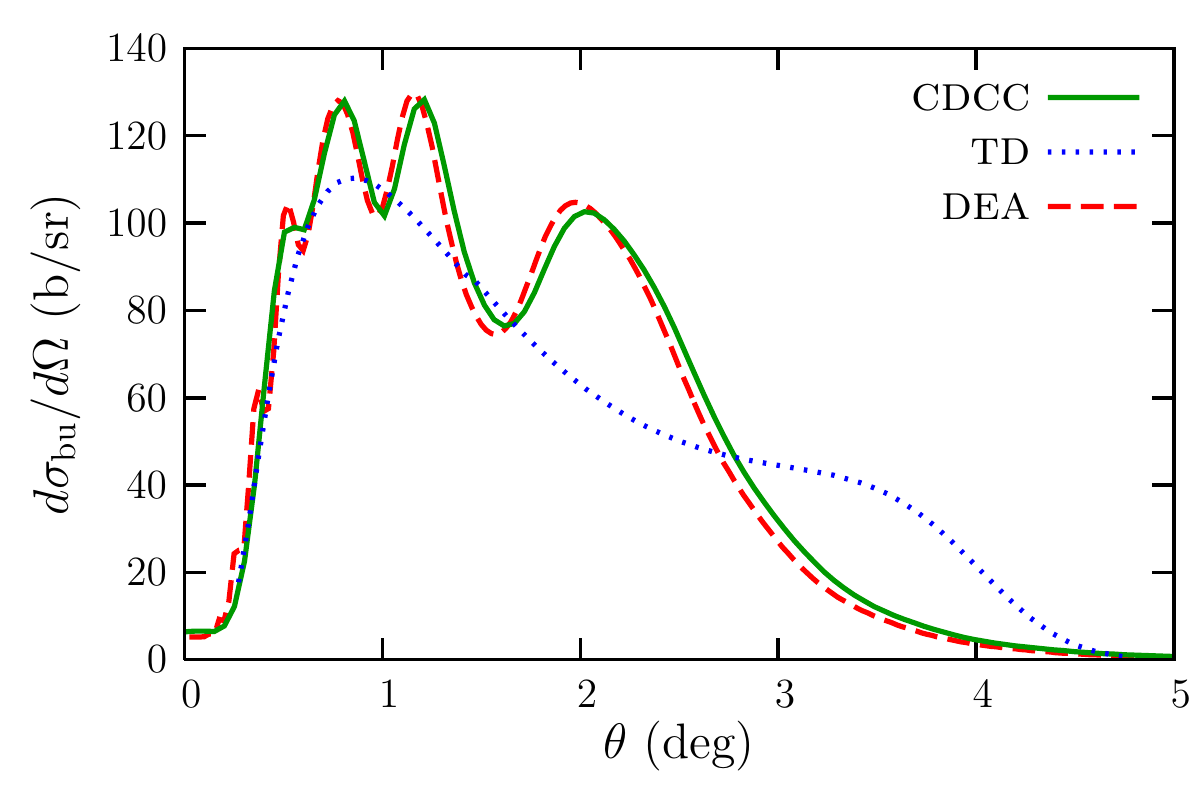}
\caption{Comparison of CDCC, DEA and TD, three nonperturbative breakup models used to analyse experiments involving halo nuclei \cite{CEN12}.
Energy distribution (left) and angular distribution (right) for \ex{15}C impinging on Pb at $68A$MeV, data from Ref.~\cite{Fuk04}.}\label{f2}
\end{figure}

In Ref.~\cite{CEN12}, these three models have been confronted to one another for \ex{15}C impinging on Pb at $68A$MeV.
All three models agree very well with each other in the energy distribution (see \fig{f2} left) indicating that they describe the same reaction process.
They also agree very well with the data of \Ref{Nak09}, confirming the halo structure of \ex{15}C.
In the angular distribution (see \fig{f2} right), CDCC and DEA agree very well with each other.
TD however, due to its underlying semiclassical approximation, does not exhibit the oscillatory pattern,
that corresponds to quantal interferences.
Nevertheless, it gives the general trend of the distribution, which explains why it reproduces correctly observables integrated over the angles such as the energy distribution.

This example illustrates the reliablity of reaction modelling.
However it is mostly confined to a simple description of the projectile.
Various groups have tried to go beyond this simple two-body description, e.g.\ by extending the aforementioned models to three-body projectiles.
A 4-body CDCC reaction models, meaning a three-body projectile plus a target, has been developed by two groups \cite{Mat04,Rod08}.
Unfortunately, due to the heavy computational requirement of CDCC, this extension was at first limited to the description of the elastic scattering of two-neutron halo nuclei, in which a breakup channel was included in a crude way.
This was nevertheless interesting as it showed that both the three-body structure of the projectile and the inclusion of the breakup channel are necessary to reproduce experimental data.

Another group suggested instead to use a simpler reaction model within which a full description of the three-body continuum could be included \cite{BCD09}.
They used the Coulomb-Corrected Eikonal model, which corresponds to an approximate solution of the DEA equation \eq{e10}.
First applied to \ex{6}He (see \Ref{BCD09} and \fig{f6}), this model showed excellent agreement with experiment for the Coulomb breakup of \ex{11}Li on Pb at $70A$MeV \cite{PDB12}, which confirmed the two-neutron halo in \ex{11}Li and enabled to study the reaction mechanism for such kind of nuclei.
This study suggests in particular that double-differential cross sections like Dalitz plots provide information about correlations between both halo neutrons (see also \Sec{C0}).

\section{Sensitivity of breakup reactions to the projectile structure}\label{structure}
\subsection{Peripherality of breakup reactions}

\begin{figure}[h]
\center
\includegraphics[trim=4cm 15.5cm 4cm 4cm, clip=true, width=8cm]{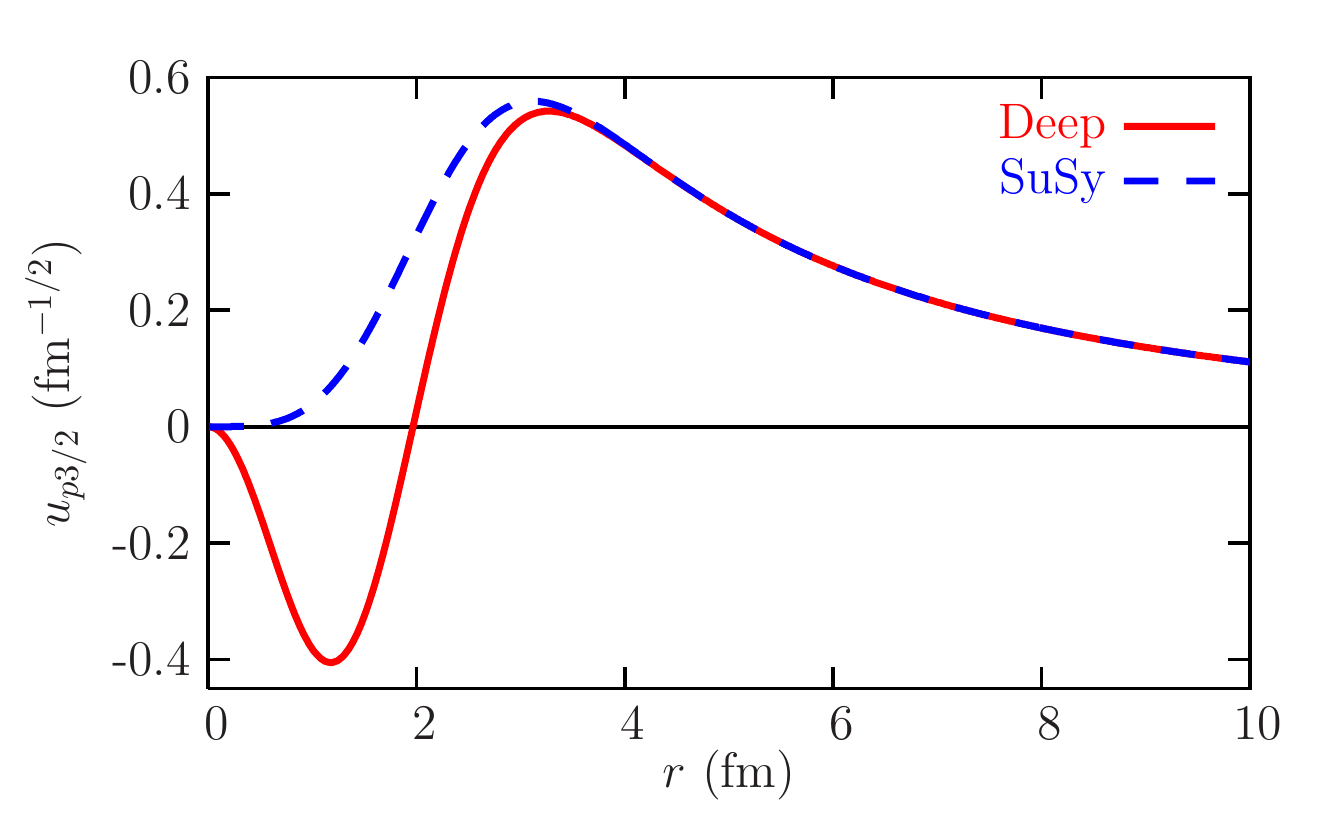}
\includegraphics[trim=2cm 1.8cm 3cm 12.5cm, clip=true, width=6.5cm]{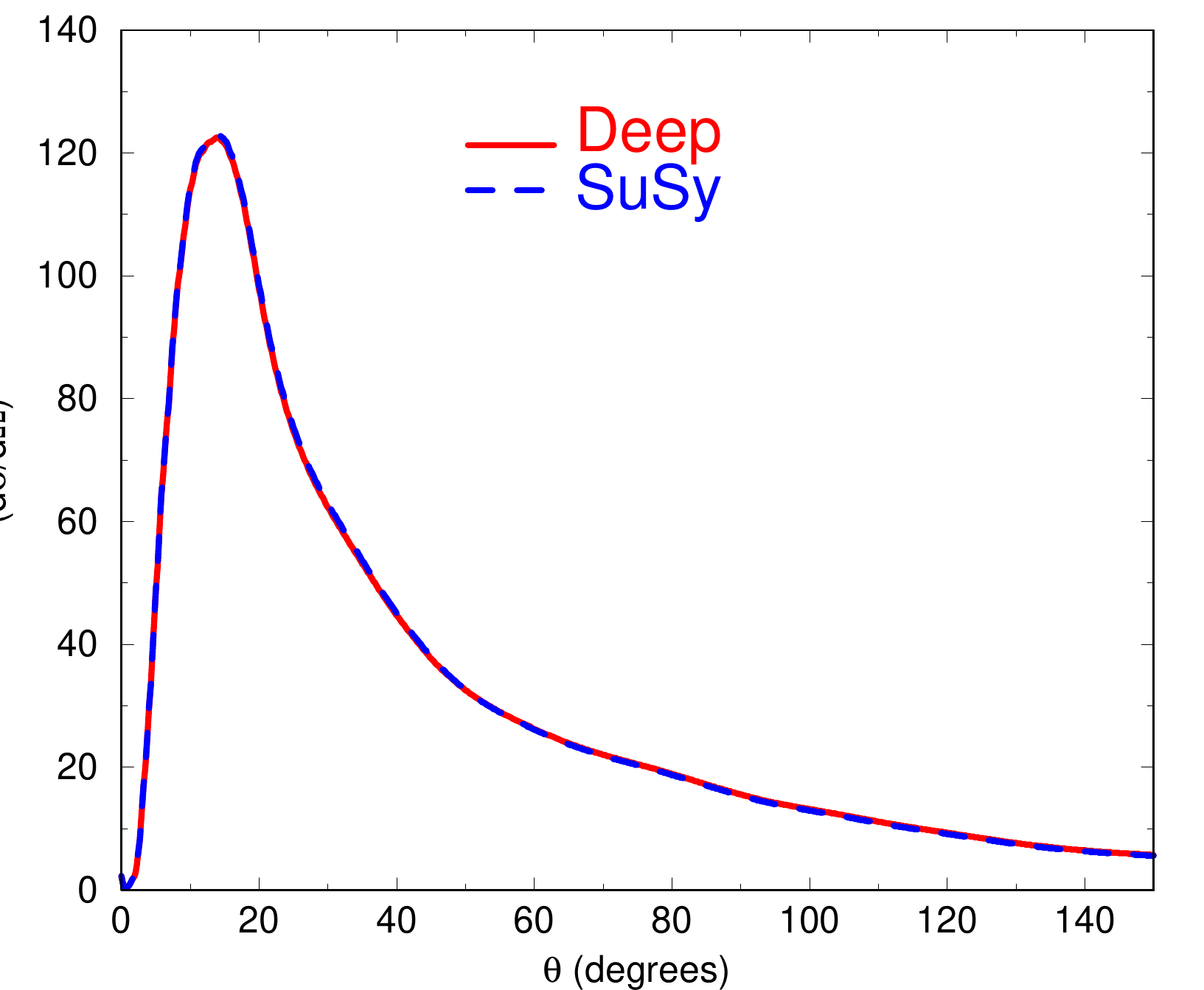}
\caption{Peripherality of breakup reactions: bound-state wave functions that differ significantly in the interior but have identical asymptotics (left) lead to identical breakup cross section (right).
CDCC calculation performed for the breakup of \ex{8}B on Ni at 25~MeV \cite{CN07}.}\label{f4}
\end{figure}

Breakup measurements are often used to infer a spectroscopic factor (SF) for one particular configuration of the projectile \cite{Fuk04}.
However, this is valid only if the reaction probes the whole range of the projectile wave function.
To check whether or not breakup reactions are sensitive to the internal part of the projectile,
calculations have been performed with descriptions of the projectile that differ significantly in the interior but exhibit identical tails \cite{CN07}.
\fig{f4} illustrates the results obtain in the case of \ex{8}B on Ni at 25~MeV.
The left panel shows the two bound-state wave functions considered in the test, while the right panel shows the corresponding angular distributions obtained within CDCC.
As is clearly shown both curves are superimposed indicating that, albeit significant, changes in the interior of the projectile wave function do not affect breakup calculations.
Similar results have been obtained at other energies, for different projectiles and targets, indicating the generality of this conclusion \cite{CN07}.
Breakup is thus a peripheral reaction in the sense that it probes only the tail of the projectile wave function, i.e. its Asymptotic Normalisation Coefficient (ANC).

\subsection{Sensitivity to resonant continuum}\label{res}
\begin{figure}
\center
\includegraphics[trim=4cm 15.5cm 4cm 4cm, clip=true, width=10cm]{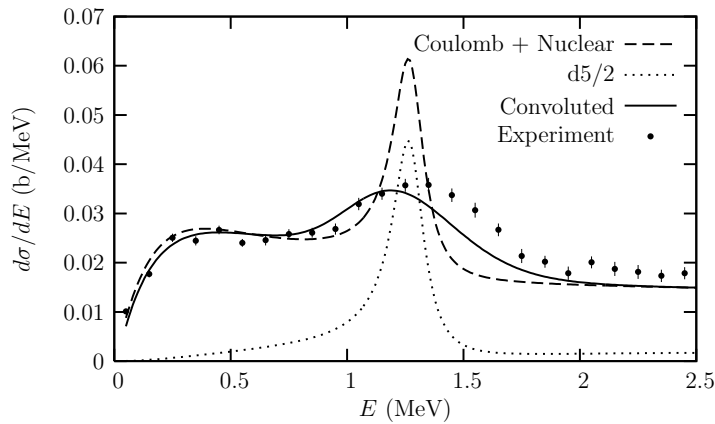}
\caption{Nuclear breakup of \ex{11}Be on C at $67A$MeV.
The calculation confirms the significant influence of the $d_{5/2}$ resonance upon the energy distribution \cite{CGB04}.
Data from Ref.~\cite{Fuk04}.}\label{f5}
\end{figure}

Along the Coulomb breakup of \ex{11}Be, Fukuda \emph{et al.} have measured its dissociation on a light target (C) (see \fig{f5}) \cite{Fuk04}.
This energy distribution exhibits a very different behaviour than Coulomb-breakup ones (compare to \fig{f2} left): it exhibits a peak at the energy of a known $5/2^+$ resonance.
This was interpreted as the signature of the \ex{10}Be-n resonant state in nuclear-dominated breakup \cite{Fuk04}.
This was confirmed by a TD calculation in which the two-body description of \ex{11}Be included this resonance within the $d5/2$ partial wave \cite{CGB04}.
The calculated cross section exhibits a narrow peak (dashed line in \fig{f5}), which, once folded with the experimental energy resolution, agrees very well with the measurements (solid line).
This peak is observed only in the $d5/2$ partial wave, confirming that nuclear breakup can be used reliably to study the presence of cluster resonant states in nuclei far from stability.

The influence of resonant states upon breakup has also been shown for three-body projectiles like \ex{6}He (see below and Refs.~\cite{BCD09,KMM13}).
This influence thus appears in more complicate structures than the simple two-body model presented here and one can expect that interesting information about the projectile continuum can be inferred from such measurements (see also \Ref{ML12}).

\subsection{Influence of the non-resonant continuum}\label{C0}
In the previous section, the influence of resonant states on breakup observables has been shown.
It is therefore necessary to properly include these states in reaction modelling.
The question addressed in this section is whether the description of the non-resonant continuum of the projectile influences breakup calculations or not.
A study has shown that for two-body projectiles, breakup cross sections can be affected by up to 40\% depending on the choice of $V_{cf}$ in the continuum \cite{CN06}.
This is significant, in particular if one is interested in extracting spectroscopic information such as ANCs from breakup data.

\begin{figure}[h]
\center
\includegraphics[trim=0cm 10cm 12cm 0cm, clip=true, width=9cm]{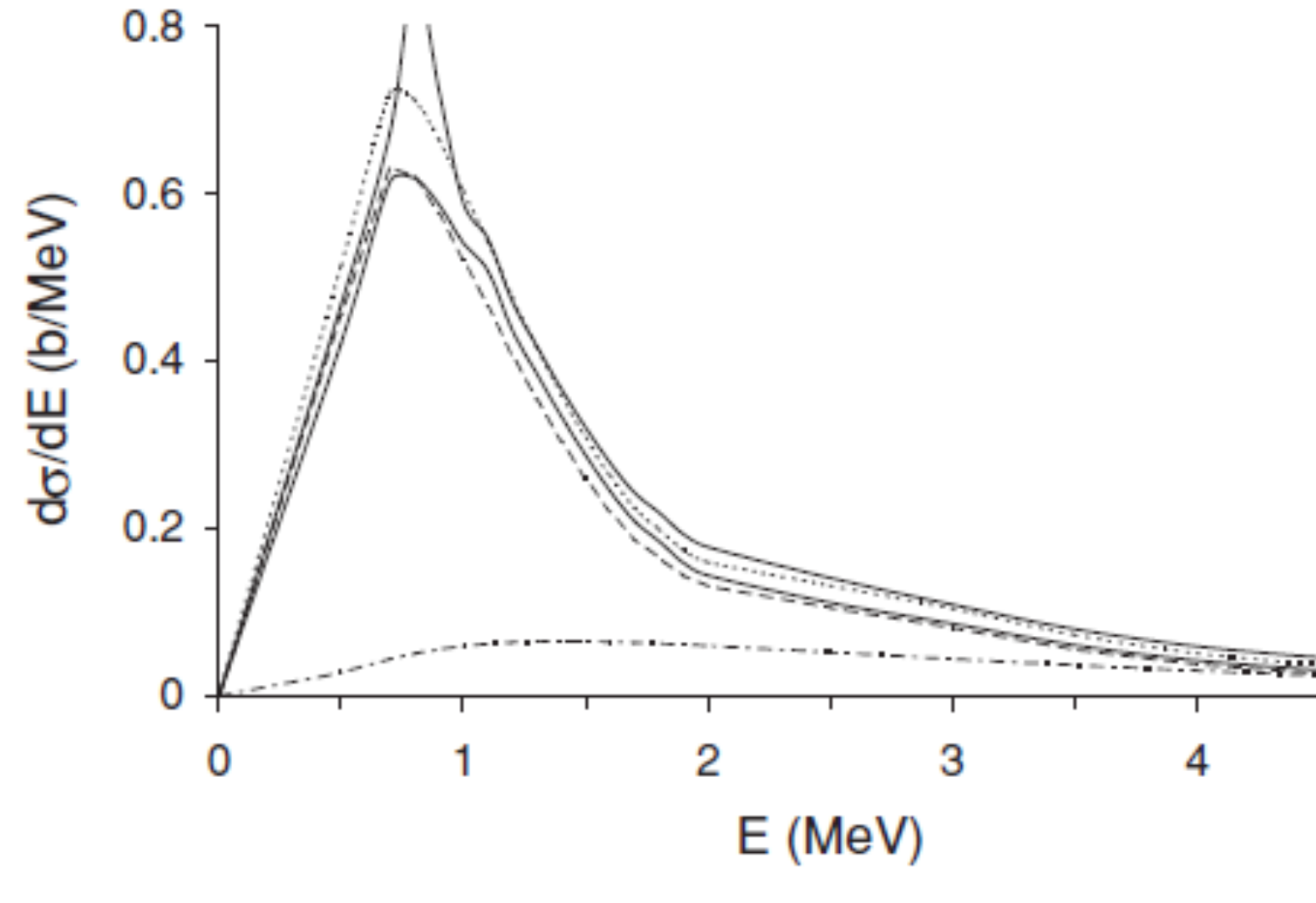}
\caption{Influence of the non-resonant continuum upon breakup calculations.
The Coulomb-breakup cross section of \ex{6}He on Pb at $240A$MeV is shown to exhibit a significant $E1$ contribution (lower solid line) when final-state interactions are considered.
This broad peak disappear when the continuum is described by plane waves (dashed-dotted line) \cite{BCD09}.}\label{f6}
\end{figure}

The problem worsens for three-body projectiles.
This is illustrated in \fig{f6}, which shows the energy distribution for the breakup of \ex{6}He on Pb at $240A$MeV, i.e. the breakup cross section as a function of the total energy between the $\alpha$ and the two halo neutrons after dissociation.
The calculation is made within the CCE \cite{BCD09} with a complete description of the continuum of \ex{6}He.
The total cross section is the top solid line, which includes the contribution of the non-resonant continuum and that of the narrow $2^+$ resonance at about 1~MeV above the $\alpha$-n-n threshold.
As already discussed in \Sec{res}, the latter is responsible for the narrow peak at that energy.
The non-resonant contribution is mostly the $E1$ strength shown by the lower solid line.
This non-resonant continuum is obtained assuming proper interactions between the clusters \cite{BCD09}.
When these interactions are switched off, i.e.\ when the continuum is described by mere plane waves, the cross section exhibits a much different behaviour (see the dash-dotted line in \fig{f6}).

The analysis of these results indicates that these final-state interactions lead to a breakup that populates a slightly dominant $\alpha$-dineutron structure in the \ex{6}He continuum.
Other theoretical works have confirmed this hypothesis of a significant effect of the final-state interactions in  breakup calculations \cite{KKM10}.
In all cases, complex structures are observed in the double-differential cross sections, which suggests that interesting information about the three-body continuum can be inferred from such observables.
In particular one might hope to study the correlation between both halo neutrons through such cross sections.
Experimental data will help us confirming these hopes.
In any case, they would enable us to discriminate the different models of \ex{6}He that have been used in these calculations.

\section{Outlook}\label{conclusion}
Breakup reactions are a powerful tool to study the structure of nuclei far from stability and in particular exotic cluster structures like halos \cite{Tan96}.
However, the reaction mechanism is more complicated than initially thought and an accurate reaction model is needed to analyse reliably experimental data.
Various such models exist; CDCC \cite{Kam86,TNT01}, TD \cite{KYS94,CBM03c} and DEA \cite{BCG05} have been reviewed in this contribution (see \Ref{BC12} for a more complete review).
They compare well to each other at intermediate energy \cite{CEN12}.
They also reproduce experimental data, which suggests that the reaction mechanism is now rather well understood, at least for two-body projectiles.

The extension of reaction models to more complicated structures, like two-neutron halo nuclei, is computationally challenging \cite{Mat06,Rod09,BCD09} but recent developments have shown it possible.
This has opened the path to an extensive study of breakup observables for three-body projectiles.

Within this contribution, the analyses of the sensitivity of breakup calculations to the projectile description have also been reviewed.
First it has been reminded that breakup reactions are peripheral in the sense that they are sensitive only to the tail of the projectile wave function and not to its whole range, i.e.\ they are better suited to extract ANCs than SFs  \cite{CN07}.
Second, they are not only sensitive to the projectile ground state, but they also probe its continuum.
In particular core-halo resonances can be studied through nuclear-dominated reactions \cite{CGB04,KMM13,Fuk04}.
Interestingly, the non-resonant continuum also influences the calculations \cite{CN06}.
This should be kept in mind when analysing experimental data, as it can affect values such as ANCs inferred from measurements.
This sensitivity is an opportunity for three-body projectiles as recent calculations have shown that their continuum has a significant influence upon breakup calculations \cite{BCD09,KKM10}.
In particular, double-differential cross sections (Dalitz plots) could be used to study the correlation between both halo neutrons in projectiles like \ex{11}Li and \ex{6}He.

Breakup reactions therefore exhibit great qualities to analyse exotic cluster structures.
Experimental and theoretical efforts should be made to refine the information that can be obtained through breakup to expand our understanding of nuclear structure far from stability.

\section*{Acknowledgments}
This research was supported in part by the Fonds de la Recherche Fondamentale
Collective (grant number 2.4604.07F).
This text presents research results of the Belgian Research Initiative on eXotic nuclei
(BRIX), Program No. P7/12, on interuniversity attraction poles of the
Belgian Federal Science Policy Office.

\section*{References}


\end{document}